\documentclass[12pt]{article}
\usepackage{graphicx}
\usepackage{color}
\usepackage{amsmath,slashed, amssymb}
\usepackage{fancyhdr}
\usepackage{hyperref}
\usepackage[titletoc]{appendix}
\numberwithin{equation}{section} 

\def\beq{\begin{eqnarray}}
\def\eeq{\end{eqnarray}}
\def\bea{\begin{eqnarray*}}
\def\eea{\end{eqnarray*}}

\def\centeron#1#2{{\setbox0=\hbox{#1}\setbox1=\hbox{#2}\ifdim
\wd1\rangle\wd0\kern.5\wd1\kern-.5\wd0\fi
\copy0\kern-.5\wd0\kern-.5\wd1\copy1\ifdim\wd0\rangle\wd1
\kern.5\wd0\kern-.5\wd1\fi}}
\def\ltap{\;\centeron{\raise.35ex\hbox{$\langle$}}{\lower.65ex\hbox{$\sim$}}\;}
\def\gtap{\;\centeron{\raise.35ex\hbox{$\rangle$}}{\lower.65ex\hbox{$\sim$}}\;}

\newcommand{\newc}{\newcommand}
\newc{\qbar}{{\overline q}}
\newc{\Kahler}{Kahler }
\newc{\deltaGS}{\delta_{\rm GS}}

\begin{document}
\begin{titlepage}
\begin{flushright}
{\large SCIPP 15/04\\UTTG-08-15 \\
}
\end{flushright}

\vskip 1.2cm

\begin{center}

{\LARGE\bf Light Scalars and the Cosmos: Nambu-Goldstone and Otherwise}

Nambu Memorial Symposium, University of Chicago, 2016

\vskip 1.4cm

{\large Michael Dine}
\\
\vskip 0.4cm
{\it $^{(a)}$Santa Cruz Institute for Particle Physics and
\\ Department of Physics, University of California at Santa Cruz \\
     Santa Cruz CA 95064  } \\
\vspace{0.3cm}

\end{center}

\vskip 4pt

\vskip 1.5cm

\begin{abstract}
This talk focuses on the role of light scalars in cosmology, both Nambu Goldstone bosons and pseudo moduli.  
The former include QCD axions, which might constitute the dark matter,  and more general axions, which, under certain conditions,
might play the role of inflatons, implementing {\it natural inflation}.  The latter are the actors in (generalized) hybrid inflation.
They rather naturally yield large field inflation, even mimicking chaotic inflation for suitable ranges of parameters.
\end{abstract}

\end{titlepage}




\section{Introduction: Homage to Nambu}

From my graduate student days, when I first encountered his work on symmetry breaking in the strong interactions, Nambu has been one of my intellectual heroes.  This was reinforced during my years at City College, where I regularly heard stories of Nambu from Bunji Sakita, who himself was an admirer.   While he was somewhat younger than Nambu, Sakita often regaled me with stories of the War years in Japan, involving both his experiences and Nambu's.  I finally got to know Nambu in the 1980's, and all of my interactions with him were intellectually stimulating and enhanced by his charm and wit.  I remember many conversations at Chicago, but remarks he made at the 1984 Argonne meeting on String Theory, which were thoughtful but cautionary, particularly stand out.  My final interactions came shortly after his Nobel Prize.  Like many, I sent him a congratulatory note, only to receive a "mailbox is full" reply.  About a year later, though, I received the most thoughtful note.

In my own work, Nambu's influence is perhaps heaviest in the areas of string theory and in the appearance of light scalars in the case of continuous global symmetry breaking.  His work with Jona-Lasinio has always been instructive for me in that it takes a model which in detail cannot be taken too seriously, but extracts important, universal features.  Some of what I say in this talk I hope can be viewed as a modest effort in this style.  While much of my discussion will center on Nambu-Goldstone bosons, I will also consider some questions in strong dynamics, where fermionic condensates will play important roles.

In his famous work on symmetry breaking, the light scalars were the pions of the strong interactions.  Today, particularly important roles for light scalars arise in cosmology.  Examples include axions as dark matter, but also candidates for the inflaton of slow roll inflation.
Indeed, the following statements are often made about inflation:
\begin{enumerate}
\item  The Planck satellite ruled out hybrid inflation
\item  If tensor fluctuations are observed (requiring Planck scale variation of the inflaton) then inflation is necessarily ``natural",
or ``chaotic".
\item  In the case of ``natural" inflation, this must be understood as ``monodromy" inflation or ``aligned axions".
\end{enumerate}
Today I want to push back (gently) on these statements.

\section{ Small vs. Large Field Inflation}

We often distinguish two ategories of Inflationary models:
\begin{enumerate}
\item   Large Field Inflation:  $\phi \gg M_p$.  In this rubric, most often one speaks of chaotic inflation or natural inflation and its
variant, monodromy inflation.  This framework predicts potentially observable gravity waves.
\item   Small Field Inflation:  $\phi \ll M_p$.  This is often associated with ``hybrid inflation".  In this case, there are no observable gravity waves.  For small
field hybrid inflation, the challenge is to understand $n_s$ as reported
by Planck ($n_s =  0.9603 \pm 0.0073.$)  Indeed, the Planck paper\cite{planck} asserts that hybrid inflation is ruled out.
\end{enumerate}


\subsection{Theoretical Challenges For Large Field Inflation}

Both the small and large field scenarios for inflation face significant theoretical challenges.
For the two principle implementations of large field:
\begin{enumerate}
\item   Chaotic inflation\cite{chaotic} is typically modeled with monomial potentials; arguably this is the  {\it definition}
of chaotic inflation.  One requires very small coefficients for the monomial, and this is typically not enforced by any conventional
symmetry (often approximate, continuous shift symmetries for non-compact fields are invoked, but we have little
understanding of how such symmetries would arise, say, in string theory).  The dominance of a particular
monomial requires even greater suppression of a host of terms of the form $\phi^n \over M_p^{n-4}$ for many $n$.  Again, why should this be?
\item   Natural inflation:  natural inflation invokes Nambu-Goldstone bosons as inflatons\cite{naturalinflation}.  This is, indeed, the sense
in which natural inflation is strictly natural according to the criterion of 't Hooft\cite{thooftnaturalness}:  such models contain small parameters protected by approximate
symmetries.
Axions typically have a periodicity (discrete
shift symmetry),
\beq
a \rightarrow 2 \pi f_a
\eeq
where $a$ is a canonically normalized scalar field.  
So $a$ is compact.  Correspondingly, the potential for $a$ is of the form:
\beq
V(a) = \Lambda^4 \cos ({a \over f_a} + \delta).
\eeq
Successful inflation requires that $a$ vary over a range much larger than the Planck mass,
i.e.  $f_a \gg M_p$.  This is in the realm of graviational physics.  The one framework in which we can assess the plausibility of such large $f_a$ is string theory,
where this doesn't seem to be realized in known constructions\cite{dineetalnatural}.  This observation has been elevated to a principle, the
{\it Weak Gravity Conjecture}, which makes natural inflation, as originally proposed, seem unlikely\cite{weakgravityconjecture}.
\end{enumerate}

In recent
years, a plausible generalization of natural inflation has emerged:  monodromy inflation\cite{silversteinwestphal}.   Theories of monodromy inflation
mimic a large $f_a$ by allowing the axion to vary over a much larger range.  This can be understood in terms of a potential:
\beq
V(a) = \Lambda^4 \cos ({{a \over f_a} + 2 \pi k \over N}).
\eeq
The $2 \pi$ periodicity survives if $k \rightarrow k-1$ when $a$ shifts.  This phenomenon was argued to be realized in string theory in \cite{silversteinwestphal}; here
we'll see some simple (and in fact familiar) realizations in field theory.  We should note that there are related proposals, such as
``multi-natural inflation", which we will not consider here\cite{choinatural}.

\subsection{Theoretical Challenges for
Small field Inflation}

Here we will focus mainly on hybrid inflation.  Hybrid inflation is often understood in terms of rather specific
(usually supersymmetric) models, but it can be understood in general terms as inflation on a pseudomoduli space.  In its
favor, as we explain, 
it is somewhat natural (arguably much moreso than chaotic inflation).  But Planck scale corrections are still important and must
be suppressed.  There is no obvious symmetry explanation for this suppression, so there appears an irreducible tuning at the 
level of $10^{-2}$ or so.  One also requires very tiny couplings to account for $\delta \rho \over \rho$.  This smallness may be technically
natural, but it is disturbing nevertheless.

More generally, there is a challenge in either the small or large field frameworks:  to what extent can one make predictions which would tie to a detailed microscopic picture.
We will propose an alternative viewpoint on modeling inflation in this talk, but we won't give a completely satisfying answer to this question.

\section{Nambu-Goldstone Bosons as the Actors in Inflation:  {\it Natural Inflation} and its Variants}

With
\beq
V(a) = \Lambda^4 \cos(\theta/f_a).
\eeq
The constraints on $f_a$ arise from satisfying the slow roll conditions, for example:
\beq
\eta =M_p^2 {V^{\prime \prime} \over V} \ll 1
\eeq
require $f_a \gg M_p$.  As we have mentioned, this is difficult to realize in string theory.  Silverstein and Westphal\cite{silversteinwestphal} suggested an alternative.
They noted that  in string models in the presence of branes, axions can ``wind".  They have, as a result, an approximate periodicity greater than $2\pi$.
Silverstein and Westphal found that the  potentials for these fields were monomial, yielding a form of chaotic inflation.

The string constructions are somewhat complicated.  A readily understood class of models of monodromy
inflation can be exhibited in field theory\cite{dinedrapermonodromyinflation}. 
Consider an SU(N) supersymmetric gauge theory without chiral matter.  In such a theory, there is a gaugino
condensate.  The idea of non-zero fermion bilinears goes back, of course, to Nambu.  In the strong interactions, these have been studied phenomenologically
since that time, and more quantitatively in lattice gauge theory.  In the last few decades, supersymmetry has provided a context in which
such condensates can sometimes be computed analytically.  In particular, in the $SU(N)$ supersymmetric gauge theory\cite{seibergholomorphy},
\beq
\langle \lambda \lambda \rangle = \Lambda^3 e^{2 \pi i k \over N} e^{i \theta/N}.
\eeq
The phase reflects the spontaneous breaking of a $Z_N$ symmetry.

If we perturb the theory with a susy-breaking gaugino mass term, $m_\lambda \lambda \lambda$, then
\beq
V(\theta) = m_\lambda \Lambda^3 \cos ({\theta \over N} + {2 \pi k \over N}).
\label{thetapotential}
\eeq
So, as we anticipated above, the naive periodicity $\theta \rightarrow \theta + 2 \pi$ is compensated by changing the branch, $k$.
If we elevate  $\theta$ to a (pseudo) Nambu-Goldstone boson, $\theta = a(x)/f_a$, then if the axion moves slowly, it simply
crosses to the other branches.  The tunneling rate scales as
\beq
\Gamma \propto e^{-N^4 \left ({\Lambda \over m_\lambda} \right )^3}.
\eeq
so the tunneling rate for changes of $k$ is extremely small\cite{shifmantunneling,dinedrapermonodromyinflation}.
Further observations on the tunneling rate will appear  in \cite{dineetaltoappear}.

For sufficiently large $N$ ($N \sim 50-100$) and $m_\lambda$ small (not too small, as will be clear from
\cite{dineetaltoappear}, we seem to have what we require for
successful natural inflation.  One might wonder whether this a  particularly plausible story (for example.  For example, is such large $N$ in the {\it Swampland}?\cite{swampland}).  Similar questions can be raised about 
the ingredients of brane constructions, which one might think roughly dual to these field theories.

\subsection{ Other Field Theory Realizations}

Witten, long ago, put forward a particularly interesting proposal for monodromy in QCD\cite{witteninstantonsnot,wittenetaprime,wittenetaprimetwo, witten1998}.  He advocated considering QCD from the viewpoint of the Large N 
expansion. In large $N$, he argued that instanton
effects should be exponentially suppressed.  Instead, he proposed that quantities like correlation functions of $F \tilde F$, even
at zero momentum, should exhibit the same behavior with $N$ as in perturbation theory.  For example, in pure gauge theory,
a correlator
\beq
\langle \left  (\int d^4 x F \tilde F \right )^n \rangle \propto N^{2-n}.
\eeq
To account for the $2\pi$ periodicity of $\theta$, he suggested that pure QCD, for example, should have $N$ branches,
with the $\theta$ dependence of the vacuum energy, for example behaving as
\beq
E(\theta) = {\rm min}_k~ (\theta + 2 \pi k)^2.
\eeq

It is interesting to revisit these questions in light of our understanding of instantons and $\theta$ in supersymmetric theories,
where we have a great deal of theoretical control.  This control extends to the inclusion of small soft breakings; of course,
we need to pass to large soft breakings if we are to understand real QCD.  In any case, already examining eqn. \ref{thetapotential},
we see exactly the sort of structure anticipated by Witten.  In fact, exploring these theories, one can also reproduce
Witten's modification of the non-linear lagrangian of the pseudogoldstone bosons to include the $\eta^\prime$.  The origin
of the branches in each of these cases is clear:  they are associated with the breaking of an approximate $Z_N$ symmetry.

But said this way, it is natural to ask what happens as one increases, say, $m_\lambda$, so that the discrete symmetry
is badly broken -- and ordinary QCD recovered.  One might speculate that the $N$ branches would collapse into a small number,
and a more conventional periodicity in $\theta$ would be recovered.  The problem with this idea is the presumed suppression of
instanton effects.

So it is is interesting to revisit these as well.  This will be done in \cite{dineetaltoappear}, but the main point is simple,
and not surprising given our experience with SUSY theories.  In cases where instanton calculations are reliable, they are not
suppressed with $N$.  In fact, they reproduce the counting expected from perturbation theory.  So on the one hand, the
instanton argument for branches does not hold; on the other hand, where one has control, Witten's conjectures are realized.
In \cite{dineetaltoappear}, possible behaviors are enumerated, and a set of possible lattice tests will be enumerated.

To summarize this section, we have considered the two popular realizations of large field inflation:
\begin{enumerate}
\item  Chaotic inflation with monimials:  Here one has the longstanding puzzle of accounting
for the suppression of an infinite set of operators -- polynomials in fields -- at large fields.  These are non-compact fields 
and it is hard to see how this arises from conventional symmetries.
\item  Monodromy inflation has simple realizations in field theory, but these require very large gauge groups.  It is not
clear how plausible this is.  Whether similar plausibility issues exist for the string constructions would seem a question
worthy of investigation.
\end{enumerate}
So it is interesting to consider other possibilities for large field inflation.  Indeed, 
string theory suggests a simple alternative picture.

\section{Non-Compact String Moduli as an Arena for Inflation}

String theory possesses non-compact fields which, at least in certain regions of the field space, would seem likely
to develop very flat potentials.  These are the {\it moduli} of the theory.
The idea that such fields are candidate inflatons has been around a long time (e.g. \cite{banksmoduliinflation}).  Here we will add some new elements.

To justify the existence of a pseudomoduli space, we assume approximate supersymmetry.   The non-compact
moduli of interest are typically scalar partners of compact moduli (axions).
Together these form the lowest
components of superfields, $\Phi = r + i a$.  The discrete shift symmetry of the axion, as well as the requirement of sensible behavior
in asymptotic regions of the moduli space (e.g. where the theory is typically weakly coupled) allow one to write,
for example, the superpotential as a series of terms of the form $e^{-N \Phi}$.  This connection
between axions and moduli will provide us with a conceptual peg for our discussion.  We will see that the
combination of
large and small field inflation has  close parallels to large and small field solutions of strong CP problem.
In this section we will:
\begin{enumerate}
\item  Review  (small field) hybrid inflation extracting some general lessons.
\item  Discuss large and small field solutions of the strong CP problem
\item  Consider Inflation with non-compact moduli
\item  Enumerate the Ingredients for successful modular inflation
\item  Discuss large field excursions in the moduli space
\end{enumerate}

\subsection{ Hybrid Inflation:  Small Field}

Hybrid inflation is often defined in terms of fields and potentials with rather detailed, special features, e.g. a so-called waterfall field\cite{hybrid1,
hybrid2,hybrid3,hybrid4}.
But hybrid inflation can be characterized in a more conceptual way.   Inflation occurs in all such models on a pseudomoduli space, in a region where supersymmetry is badly broken (possibly by a larger amount than in the present universe) and the potential is slowly varying\cite{dinehybrid}.

Essentially all hybrid models in the literature are small field models; this allows quite explicit constructions using rules of conventional effective field theory, but it is not clear that small field inflation is selected by any deeper principle.  
The simplest (supersymmetric) hybrid model involves two fields, $I$ and $\phi$, with superpotential:
\beq
W = I (\kappa \phi^2 - \mu^2).
\eeq
$\phi$ is known as the waterfall field.  

Classically, for large $I$, the potential is independent of $I$;
$$V_{cl} =  \mu^4 ~(\phi=0).$$
The quantum mechanical corrections control the dynamics of the inflaton:
 \beq
 V(I) =\mu^4(1 + {\kappa^2 \over 16 \pi^2} \log(\vert I \vert^2/\mu^2)).
 \eeq 
 $\kappa$ is constrained to be extremely small in order that the fluctuation spectrum be of the correct size; $\kappa$ is proportional, in fact, to $V_I$, the energy during inflation.   The quantum corrections determine the slow roll parameters.
One has:
\beq
V_I = 2.5 \times 10^{-8} \epsilon^2 M_p^4
\eeq
\beq
\kappa = 0.17 \times \left ({\mu \over 10^{15} {\rm GeV}} \right )^2 = 7.1 \times 10^5  \times \left ({\mu \over M_P} \right )^2.
\eeq

\subsubsection{Corrections to the Simplest Model}

In addition to the quantum corrections we have described, higher dimension operators of various types, even if Planck suppressed,
can have dramatic effects.  These include:
\begin{enumerate}
\item  Kahler Potential Corrections:
One expects corrections to the Kahler potential; we will assume here that they
are Planck suppressed; the constraints are more severe if they are suppressed by some smaller
scale.  We organize the effective field theory in powers of $I$\cite{dinehybrid}.   The quartic term in $K$,
\beq  
K = {\alpha \over M_p^2 }I^\dagger I I^\dagger I
\eeq
gives too large an $\eta$ unless $\alpha \sim 10^{-2}$.
This appears to be one irreducible source of fine tuning in this framework.
\item  Superpotential Corrections:
thesel are potentially very problematic.  For example, one can write:
\beq
\delta W = {I^n \over M_p^{n-3}}
\eeq
At least the low $n$ terms must be suppressed.
This might occur as a result of discrete symmetries.The leading power of $I$ in the superpotential controls the scale of inflation.
For example, $N=4$,  gives $\mu \approx 10^{11}$ GeV and $\kappa \approx 10^{-10}$.
With $N=5$, one obtains $\mu \approx 10^{13}$ GeV, and $\kappa \approx 10^{-5}$
The scale $\mu$ grows slowly with $N$, reaching $10^{14}$ GeV  at $N=7$ and $10^{15}$ GeV for $N=12$.
\end{enumerate}

This result is interesting from the perspective of understanding (predicting?) the scale
of inflation.  It is hard to understand a high scale of inflation in this framework without a rather absurd sort of discrete symmetry.  This
might be taken as an argument for a low scale of inflation.
On the other hand, pointing in the opposite direction is $\kappa$, which gets smaller rapidly with $V_0$,
In addition, achieving $n_s < 1$, consistent with Planck, required a balancing of Kahler and superpotential corrections.
Indeed, the abstract of the Planck theory paper\cite{planck} includes the assertion:  ``the simplest hybrid inflationary models, and monomial potential
models of degree $n>2$ do not provide a good fit to the data."

Because of the requirement of small parameters, and the inevitably significant amount of fine tuning, the theoretical arguments for small field models over large field models for inflation are hardly so persuasive.
Even at low scales it is necessary to have control over Planck scale corrections, and tuning of parameters (at least at the part in $10^{-2}$ level) is required.  One also  needs a very small dimensionless parameter, progressively smaller as the scale of inflation becomes smaller.   Quite likely, any successful model requires significant discrete symmetries (or even more severe tunings).

\subsection
{Generalizing hybrid inflation to large fields:  moduli inflation}

So it is clearly interesting to explore the possibility of inflation on (non-compact) moduli spaces with fields undergoing variations of order Planck scale or larger.  Such moduli spaces are quite familiar from string theory.  
First it is instructive to consider another situation where such a small field/large field dichotomy arises:  the axion solution to the 
strong CP problem.

\subsubsection{Small Field and Large Field Solutions to the Strong CP Problem}

To {\it solve} the strong CP problem one must account for an accidental global symmetry which is of extremely high {\it quality} .
Most models designed to obtain a Peccei-Quinn symmetry can be described as small field models; they are constructed with small axion decay constant, $f_a \ll M_p$, with
$f_a = \langle\phi\rangle$  In this case, one can organize the effective field theory in powers of $\phi/M_p$ (again, if higher dimension
operators are suppressed by a scale $M \ll M_p$, the objections discussed here to the Peccei-Quinn solution are even more
severe).

We can define a notion of Axion {\it Quality}\cite{dineetalquality}.
We require
$$Q_a \equiv {1 \over f_a m_a^2}{\partial V \over \partial a} = 10^{4} {f_a} {\partial V \over \partial a} < 10^{-11}.$$

In small field models, if the axion is the phase of $\phi$, the PQ symmetry is the transformation $\phi \rightarrow e^{i \alpha} \phi$.
This symmetry must be {\it extremely} good\cite{kamionkowski}.  One
needs to suppress $\phi^N \over M_p^{N-3}$ up to very high $N$.  E.g. $Z_N$, with $N> 11$ or more, depending on $f_a$.
This is not terribly plausible.  It involves models of a  high degree of complexity, designed to solve a problem
of essentially no consequence (small $\theta$ is not, by itself, singled out by anthropic or similar considerations\cite{dineetalquality,ubalditheta}.

\subsubsection{Large field solutions of the Strong CP Problem}

String theory has long suggested a large field perspective on the axion problem\cite{wittenso10}. String theory,
as we have stressed, frequently possesses axions.  These axions exhibit continuous shift symmetries in some approximation (e.g. perturbatively in the string coupling).  Non-perturbatively these symmetries are broken, but usually one has a discrete shift symmetry left which is exact:
\beq  a \rightarrow a + 2 \pi.\eeq
We have normalized $a$ to be dimensionless; $f_a$ depends on the precise form of the axion kinetic term.  The (non-compact) moduli which accompany these axions typically have Planck scale vev's. 
Calling the full chiral axion superfield ${\cal A} = s + ia + \dots$,
this periodicity implies that, for large $s$, in the superpotential the axion appears as $e^{-\cal A}$.  Solving
the strong CP problem then requires suppressing only a small number of possible terms\cite{bobkovraby,dinefestucciawu}.

As always, in string theory, one has to understand stabilization of moduli. More honestly (thinking of Nambu's cautionary remarks)
in the current state of our knowledge, we can at best conjecture that moduli are stabilized.  If string theory is to produce an axion which can
solve the strong CP problem, typically several moduli must be stabilized.  Whatever the mechanism, the axion multiplet is special.  If the superpotential plays a significant role in stabilization of the {\it saxion}, it is difficult to understand why the axion should be light. $e^{-{\cal A}}$ would badly break the PQ symmetry if responsible for saxion stabilization.
So the stabilization must result from Kahler potential effects (presumably connected with supersymmetry breaking).
 In perturbative string models
the Kahler potential is often a function of ${\cal A} + {\cal A}^\dagger$.  There is no guarantee that would-be corrections to $K$ which stabilize ${\cal A}$ do not violate this symmetry substantially, but will take as a hypothesis.  

For example, as a model, suppose there is some other modulus, $T = t + ib$, appearing in the superpotential as $e^{-T}$, where $e^{-T}$ might set the scale for supersymmetry breaking.  
\beq
W(T) = A e^{-T/b} + W_0
\eeq
with small $W_0$, leading to
\beq
T \approx b \log(W_0).
\eeq
The potential for $s$ would arise from terms in the supergravity potential:
\beq
V_s = e^{K} \left \vert {\partial K \over \partial {\cal A}} W \right \vert^2 g^{{\cal A}~{\cal A}^*} + \dots
\eeq
For suitable $K({\cal A},{\cal A}^*)$, $V$ might exhibit a minimum as a function of $s$.    If $s$ is, say, twice $t$ at the minimum, $e^{-{\cal A}}$
is severely suppressed, as is the potential for the (QCD) axion, $a$. 

\subsubsection{A Remark on Distances in the Modulus Geometry}

Typical metrics for non-compact moduli fall off as powers of the field for large field.
Defining $s$ to be dimension one, 
\beq
g_{{\cal A},{\cal A}^*}=C^2 M_p^2/s^2
\eeq
for some constant, $C$.  So large $s$ is far away (a distance of order $a~M_p \log(s/M_p)$) in field space.  If, for example, the smallness of $e^{-(s + ia)}$ is to account for an axion mass small enough to solve the strong CP problem, we might require $s \sim 110~M_p$ , corresponding to a distance of order $8 M_p$ from $s = M_p$ if $C = \sqrt{3}$.

\subsection{ Non-Compact Moduli as Inflatons}

So the strong CP problem points to Planck scale regions of field space as the arena for phenomenology.  This has parallels
in inflation.  Moduli of the sort required for the axion solution might also play a role as inflatons.
There are some plausible ingredients for moduli as the players in inflation:
\begin{enumerate}
\item  In the present epoch, one or more moduli which are responsible for hierarchical supersymmetry breaking.
\item  In the present epoch, a modulus whose superpotential is highly suppressed, and whose compact component is the QCD axion.  This is not necessary for inflation, but is the essence of a modular (large field) solution to the strong CP problem.
\item  At an earlier epoch, a stationary point in the effective action with higher scale supersymmetry breaking then
at present and a positive cosmological constant.  
\item  At an earlier epoch, a field with a particularly flat potential which is a candidate for slow roll inflation.
\end{enumerate}

Fields need not play the same role in the inflationary era that they do now.  The Peccei-Quinn symmetry might be badly broken
during inflation.  Then the axion will be heavy during this period and isocurvature fluctuations may not be an issue\cite{dineanisimov}.
In such a case the initial axion misalignment angle, $\theta_0$, would be fixed rather than being a random variable.
We know that the scale of inflation is well below $M_p^4$.  So it is plausible that even during inflation
moduli have large vev's, $e^{-{\cal A}},~e^{-T} \ll 1$, though much smaller than at present.
For example, suppose that there exist
a pair of moduli, ${\cal A}, T$ responsible for supersymmetry breaking, and an additional field, $I$, which will play the role of the inflaton.  During inflation, 
\beq
H_I \sim W \sim e^{-t}
\eeq
For typical Kahler potentials, the curvature of the $t$ and $i$ potentials will be of order $H_I$ (for $i$, this is the usual ``$\eta$ problem").  We will exhibit a model with lower curvature below.

A successful model  requires a complicated interplay between effects due to the Kahler potential and superpotential\cite{dinelargefield}. 
\begin{itemize}
\item  The {\it potential} must possess local, supersymmetry breaking minima in ${\cal A}$ and $T$, one of higher, one of lower, energy.  The former is the setting for the inflationary phase; the latter for the current, nearly Minkowski, universe.
\item  In the inflationary domain,
the potential for $I$, must be very flat over some range.
\item  In the inflationary domain, the imaginary parts of ${\cal A}$ and $T$ should have masses comparable to $H_I$ (or slightly larger), if the system is to avoid difficulties with isocurvature fluctuations.  This would arise if $e^{-s} \approx e^{-t}$.
\item  In the present universe, the imaginary part of ${\cal A}$ should be quite light and that of $I$ much lighter.
\item  There are additional constraints from the requirement that inflation ends.  For some value of Re $I$, the inflationary minimum for $T$ and ${\cal A}$ must be destabilized (presumably due to Kahler potential couplings of $I$ to ${\cal A}$ and $T$).  At this point, the system must transit to another local minimum of the potential, with nearly vanishing cosmological constant.  
\item  The process of transiting from the inflationary region of the moduli space to the present day one is subject to serious constraints.  Even assuming that there is a path from the inflationary regime to the present one, the system is subject to the well-known concerns about moduli in the early universe\cite{bankskaplannelson,ibanez}.  If they are sufficiently massive (as might be expected given current constraints on supersymmetric particles), they may reheat the universe to nucleosynthesis temperatures, avoiding the standard cosmological moduli problem.  $T$ and ${\cal A}$ are vulnerable to the moduli overshoot problem\cite{brusteinsteinhardt}, for which various solutions have been proposed.
\end{itemize}







\subsubsection
{Inflationary Models:  Large $r$}

Here we describe a simple model which yields large $r$ and satisfies some of the conditions enumerated above (because it involves
only a single field and we specify the potential only in a limited range it cannot satisfy all)\cite{dinelargefield}  For the Kahler potential we take:
\beq
K = -{\cal N}^2 \log(I + I^*).
\eeq
With
\beq
I = e^{\phi/{\cal N}}
\eeq
the kinetic term for $\phi$ is simply $\vert\partial \phi \vert^2$. 
\beq
V(\phi) = e^{-{\cal N} \phi} V_0,
\eeq
$V_0$ being the minimum of the $S$, $T$ potential.

The slow roll parameters are:
\beq
\epsilon = {1 \over 2} {\cal N}^2;~ \eta = {\cal N}^2 = 2 \epsilon.
\eeq
Note
\beq
n-1 = -2 \epsilon.
\eeq
If $r = 0.2=16 \epsilon$,\beq
n_s -1 = 0.025
\eeq
on the high end of the range favored by the Planck measurement.

This model is discussed in the Planck theory paper which rules it out based on their measurement of $n_s$.  Suitable modifications are discussed in \cite{dinelargefield}.  A model with similar features (with $\cosh$ rather than exponential potential) has been discussed in\cite{nojiri}.

\subsubsection{ Connection to Chaotic Inflation}

Chaotic inflation has, for decades, provided a simple model for slow roll inflation, and its prediction of transplanckian field motion and observable gravitational radiation is compatible with our discussion of large field modular inflation.  As we look at the moduli inflation model of the previous section (and more generally moduli models of large field inflation), we see, in fact, a realization of the ideas of chaotic inflation\cite{dinelargefield}.  Again, the potential behaves as
\beq
V \sim H_I^2 M_p^2 e^{{\cal N} \phi}
\eeq

The exponent changes, during inflation, by a factor of about $3/2$.  So we can make a crude approximation, expanding the exponent and keeping only a few terms.  If we focus on each monomial in the expansion, the coefficient of 
$\phi^p$, in Planck units, is:
\beq
\lambda^p = {10^{-8} {\cal N}^p \over p!}.
\eeq
where $N$ is the humber of $e$-folds.

We can compare this with the required coefficients of chaotic inflation driven by a monomial potential, $\phi^p$.  In this case,
\beq
\lambda_p = {3 \times 10^{-7}  \over (2Np)^{{p \over 2} - 1}}
\eeq
These coefficients are not so different.  For example, for $p=1$, the moduli coefficient is about $2 \times 10^{-9}$, while for the chaotic case it is about four times smaller; the discrepancy is about a factor of two larger for $p=2$.  So we see that these numbers, which would one hardly expect to be identical, are in a similar ballpark.

So moduli inflation provides a rationale for the effective field theories of chaotic inflation. The typical potential is not a monomial, but one has motion on a non-compact field space, over distances of several $M_p$, with a scale, in Planck units, roughly that expected for chaotic inflation.  The structure is enforced by supersymmetry and discrete shift symmetries.

\section{ Summary}

Physicists are likely a long way from writing down {\it the} microscopic model which describes inflation.  For the time being the most sensible
approach is to consider classes of models, the constraints coming from observations, and possible general features and predictions.
Here we have discussed some features of several classes of models:
\begin{enumerate}
\item  Chaotic inflation:  we have reviewed why it is puzzling as usually formulated, and have seen that something like it may arise in frameworks which
are more natural.
\item  Natural inflation in its simplest formulation unlikely.
\item  Monodromy inflation:  has simple field theory realizations, with large amounts of inflation requiring {\it very} large gauge groups; we have
speculated about the implications for the plausibility of the mechanism.
\item  Hybrid inflation:  we have stressed that hybrid inflation should be thought of as inflation on a pseudomoduli space.  from this
vantage point, large field seems plausible (the simplest forms of small field are ruled out, and even these are highly
tuned).  Large field hybrid inflation favors higher scales for inflation, and has features which can mimic chaotic inflation.
\end{enumerate}
In all cases, detailed implementations are challenging; one wants to ask whether there are any generic features one can extract ($r$, non-gaussianity,...) and compare with data.

Explaining inflation from an underlying microscopic theory is an extremely challenging problem, quite possibly inaccessible to our current theoretical technologies.  As we have reviewed, even in so-called small field inflation, it requires control over Planck scale phenomena.  Within string theory, this requires understanding of supersymmetry breaking (whether large or small) and fixing of moduli in the present universe as well as at much earlier times.  It requires an understanding of cosmological singularities, and almost certainly of something like a landscape.

We have stressed a parallel between small/large field inflation and small/large field solutions to the strong CP problem.  The existence of moduli in string models is strongly suggestive of the large field solutions to both problems.  The proposal we have put forward here is similar to the large field solutions of the strong CP problem.

Several moduli likely play a role in inflation in order to achieve the needed degree of supersymmetry breaking and slow roll.  We have noted that small $r$ is more tuned than large $r$, giving some weight to the former possibility.  We have noted the contrast with small field inflation, where extreme tuning to achieve low scale inflation is replaced by the requirement of an extremely small dimensionless coupling.

Returning to the strong CP problem, any would-be Peccei-Quinn symmetry is an accident, and the accident which holds in the current configuration of the universe need not hold during inflation; this would resolve the axion isocurvature problem.  It would imply that $\theta_0$ is not a random variable.

The inflationary paradigm is highly successful; the question is whether we can provide some compelling microscopic framework and whether it is testable.  In the present proposal, one does not attempt (at least for now) a detailed microscopic understanding, but considers a class of theories.  Within those considered here:
\begin{enumerate}
\item   Higher scales of inflation are preferred
\item   High scale axions are likely, and the idea of an {\it axiverse} gains additional plausability\cite{axiverse}.
\end{enumerate}
In a more detailed picture, one might hope to connect some lower energy phenomenon, such as supersymmetry breaking, with
inflation.

\vskip .2cm
\noindent
\noindent
{\bf Acknowledgements:}   I thank my collaborators Patrick Draper, Guido Festuccia, Laurel Stephenson-Haskins, and Lorenzo Ubaldi for the many insights they have shared with me. This work was supported in part by the U.S. Department of Energy grant number DE-FG02-04ER41286.

\bibliography{nambu_talk}{}
\bibliographystyle{utphys}

\end{document}